\documentclass[11pt, twopage, reprint, superscriptaddress, preprintnumbers,nofootinbib]{article}
\usepackage{amsmath}
\usepackage{amsmath}
\usepackage{amssymb}
\usepackage{physics}
\usepackage{cite}
\usepackage{tikz}
\usepackage[compat=1.1.0]{tikz-feynman}
\tikzfeynmanset{warn luatex=false} 

\usepackage{bookman}
\usepackage[T1]{fontenc} 
\usepackage{microtype}
\usepackage[english]{babel}
\usepackage[hmarginratio=1:1,top=32mm,columnsep=20pt]{geometry}
\usepackage[hang, small,labelfont=bf,up,textfont=it,up]{caption}

\usepackage{abstract}

\usepackage{titlesec}
\renewcommand\thesection{\Roman{section}}
\renewcommand\thesubsection{\roman{subsection}}
\titleformat{\section}[block]{\large\scshape\centering}{\thesection.}{1em}{} 
\titleformat{\subsection}[block]{\large}{\thesubsection.}{1em}{}

\usepackage{titling} 

\usepackage{hyperref}
\hypersetup{
    colorlinks=true,
    linkcolor=black,
    filecolor=magenta,      
    urlcolor=blue,
    citecolor=blue,
    pdftitle={Reparametrisation invariance, anomalies and pseudoscalar states in massless quantum electrodynamics}
    }

\usepackage{amsmath}
\usepackage{amssymb}
\usepackage{slashed}

\title{Reparametrisation invariance, anomalies and \\ pseudoscalar states in massless quantum electrodynamics}
\author{\textsc{Joshua Cesca\thanks{\texttt{\href{mailto:joshua.cesca@sydney.edu.au}{joshua.cesca@sydney.edu.au}}}, Archil Kobakhidze\thanks{\texttt{\href{mailto:archil.kobakhidze@sydney.edu.au}{archil.kobakhidze@sydney.edu.au}}}\, and Elden Loomes\thanks{\texttt{\href{mailto:elden.loomes@sydney.edu.au}{elden.loomes@sydney.edu.au}}}} 
\vspace{0.2cm} \\
\normalsize \itshape
Sydney Consortium for Particle Physics and Cosmology, \\
\normalsize  \itshape
School of Physics, The University of Sydney, NSW 2006, Australia \\
}
\date{}
\renewcommand{\maketitlehookd}{%

\begin{abstract}
\noindent 
In electrodynamics, and more generally in gauge theories, quantisation introduces an ambiguity whereby the vector current, the axial current, or both may acquire an anomalous divergence. These anomalies can be classified as relevant, conventionally associated with the infrared properties of the theory, and irrelevant, arising from the freedom in the choice of renormalisation scheme. We show that, under a axial reparametrisation of the fermion fields, the global axial symmetry can be promoted to a gauge redundancy. Requiring this gauge symmetry to be preserved at the quantum level eliminates the irrelevant anomalies, leaving a non-propagating pseudoscalar field that acts as a Lagrange multiplier enforcing the familiar axial anomaly. We further argue that, on topologically non-trivial spacetimes, in particular those with non-trivial second cohomology, this axial gauge symmetry can undergo a Higgs mechanism, rendering the pseudoscalar field an emergent propagating degree of freedom. Finally, we briefly point out an analogous emergent state associated with the conformal anomaly.
\end{abstract}
}

\begin{document}
\maketitle

\section{Introduction}

The quantisation of gauge theories involves an ambiguity that was first highlighted in the pioneering studies of the axial anomaly \cite{Adler:1969gk}. In massless electrodynamics, there exist two classically conserved currents, corresponding to the vector and axial-vector symmetries. However, upon quantisation, the regularisation of ultraviolet divergences cannot preserve the conservation of both currents. Physical consistency, and in particular gauge invariance, requires that the vector current remains conserved, while the axial-vector current develops an anomalous divergence, leading to the violation of the axial charge. This mechanism provides a remarkable explanation of the $\pi\to\gamma\gamma$ decay process \cite{Bell:1969ts, Adler:1969gk}, which would otherwise be forbidden by axial symmetry. This anomaly also means that one cannot simultaneously gauge both symmetries coupled to arbitrary gauge field backgrounds, an obstruction known as 't Hooft anomaly. 

Conventionally, anomalies are distinguished as relevant and irrelevant. The former are associated with the infrared properties of the theory, whereas the latter arise from ambiguities in the choice of renormalisation scheme. Consequently, irrelevant anomalies can be removed by an appropriate choice of renormalisation prescription. We note, however, that although such scheme-dependent irrelevant anomalies may manifest themselves symmetrically as anomalous divergences of both axial and vector currents, the gauge invariance may remain intact, thereby ensuring the consistency of the quantised electromagnetic theory. Therefore, if one adopts the viewpoint that different regularisation schemes represent a genuine non-uniqueness of the quantum theory corresponding to a given classical theory, the selection of the physically realised quantum theory appears to be determined by empirical input rather than by theoretical consistency alone.

In quantum electrodynamics (${\rm QED}$) with massless fermions, the infrared manifestation of the axial anomaly appears as a massless pole in amplitudes involving anomalous processes, first identified in Ref.~\cite{Dolgov:1971ri}. This pole corresponds to the exchange of a massless pseudoscalar collective excitation composed of collinear massless fermion--antifermion pairs. However, this collective mode does not correspond to an asymptotic particle state, consistent with the fact that the number of propagating degrees of freedom in the quantum theory remains identical to that of the underlying classical theory. The infrared behaviour of massless electrodynamics is further complicated by the fact that the zero-momentum limit corresponds to a conformal fixed point of the theory\footnote{Throughout this paper, we assume that the theory is understood as an effective theory with a cutoff $\Lambda$ below the ultraviolet Landau-pole scale. In fact, the conformal fixed point may be viewed as an infrared counterpart of the ultraviolet Landau-pole problem \cite{Landau:1954cxv}, corresponding to the limit $\Lambda\to\infty$ with finite $m$ held fixed.}, where the notion of asymptotic particle states becomes subtle. In particular, it has been argued that massless charged fermions do not exist as asymptotic particle states \cite{Case:1962zz, Gribov:1981jw}. Instead, they are realised as infraparticles \cite{Schroer:1963gw}: charged fermionic states dressed by a soft-photon cloud \cite{Kulish:1970ut}. As a consequence, their propagators lack an isolated particle pole and instead exhibit a branch cut associated with a continuum of soft-photon excitations. At the same time, the trace (conformal) anomaly of massless electrodynamics also manifests itself through an infrared pole in the corresponding amplitudes \cite{Giannotti:2008cv}, raising the possibility of an intriguing interplay between conformal and axial anomalies. 

In this work, we treat the theory of massless $\rm QED$ with a cutoff, where quantisation ambiguities result in not necessarily equivalent effective theories. In particular, we argue that the removal of irrelevant anomalies can be traced
to the invariance of the theory under axial reparametrisations of the fermion fields; the axial reparametrisation
\begin{equation}
\psi = e^{i\gamma_5\sigma}\chi\,, \qquad
\bar{\psi} = \bar{\chi}\,e^{i\gamma_5\sigma}\,,
\label{rep}
\end{equation}
introduces a gauge redundancy under the local transformations
\begin{equation}
\chi' = { e}^{i\gamma_5\alpha}\chi\,,\quad \bar\chi'=\bar\chi\,{ e}^{i\gamma_5\alpha}\,,\quad
\sigma' = \sigma - \alpha\,.
\label{gauge}
\end{equation}
This is equivalent to the axial gauge transformation of the fermion fields coupled to an axial gauge field in the vacuum, $A_\mu=\partial_\mu \sigma$. Although gauging the axial symmetry on general $A_\mu$ backgrounds is obstructed by the 't Hooft anomaly, this restricted gauging is sufficient to eliminate the irrelevant anomalies without spoiling the theory.
The invariance of the theory under these transformations implies that the pseudoscalar field $\sigma$ does not represent an independent physical degree of freedom. In other words, the theory formulated in the original fields $(\bar\psi, \psi, V_{\mu})$ is equivalent to the theory formulated using $(\bar\chi, \chi, V_{\mu}, \sigma)$. We show that demanding the gauge symmetry (\ref{gauge}) to be preserved upon quantisation eliminates the irrelevant anomaly and renders $\sigma$ a non-dynamical Lagrange multiplier that enforces the familiar axial anomaly. Conversely, in the presence of an irrelevant anomaly, the gauge symmetry (\ref{gauge}) is no longer operative, rendering $\sigma$ a dynamical degree of freedom that is absent in the classical theory.\footnote{In this work, we focus on vector-like ${\rm QED}$. In the case of the axial ${\rm QED}$, it is well known that the consistent quantisation requires the introduction of counterterms that restore gauge invariance \cite{Krasnikov:1985bn, Faddeev:1986pc}. These counterterms place the theory in the Higgs regime \cite{Preskill:1990fr}, necessitating an additional longitudinal degree of freedom associated with the would-be Goldstone boson.}

We next argue that the physical picture changes dramatically when the spacetime manifold is topologically non-trivial, possessing a non-vanishing second cohomology, as is the case, for example, in black-hole spacetimes. The resulting topological structure of the electromagnetic gauge theory \cite{Deser:1980kc, Isham:1983qy,  Kobakhidze:2026pwp} gives rise to both a non-vanishing topological susceptibility and normalisable fermion zero modes \cite{Arunasalam:2018eaz, new}. The latter trigger spontaneous axial symmetry breaking through the formation of a fermion condensate \cite{Dvali:2024dlb,Dvali:2024zpc,Dvali:2025pcx}, while the former ensures that the symmetry breaking is realised in the Higgs phase \cite{Dvali:2005an,Dvali:2005ws, Aurilia:1980jz}. As a consequence, the pseudoscalar field $\sigma$, previously associated with the non-dynamical anomaly pole of the axial anomaly \cite{Dolgov:1971ri}, becomes a massive propagating degree of freedom. This provides another example of the general phenomenon whereby topologically non-trivial gauge theories, in the presence of fermion zero modes and the associated condensates, give rise to emergent scalar degrees of freedom \cite{Dvali:2024dlb,Dvali:2024zpc,Dvali:2025pcx}.    

The rest of the paper is organised as follows. In the next section we analyse the anomaly diagrams and argue that the gauge symmetry associated with the redundancy of the axial reparametrisation (\ref{gauge}) leads to the removal of irrelevant anomalies. In Sec.~\ref{sec3} we overview topological features of electromagnetic gauge fields on topologically non-trivial manifolds and argue for a non-perturbative fermion condensates that trigger the Higgs mechanism for the axial reparametrisation gauge symmetry and lead to an emergent axion state. The final section is reserved for conclusions and further discussions.

\section{Gauge redundancy of axial reparametrisation and relevant and irrelevant anomalies} 
\label{sec2}
We begin with the path integral for the fermion sector of ${\rm QED}$:
\begin{equation}
Z_{\rm QED}[V_{\mu}]= \int D\bar\psi D\psi \,e^{iS_{\rm QED}[\bar\psi,\psi, V_{\mu}]}\,,\label{QED_Z}
\end{equation}
where
\begin{equation}
S_{\rm QED}[\bar\psi,\psi, V_{\mu}]=\int d^4x\,\bar{\psi}i\gamma^\mu(\partial_\mu - iV_\mu)\psi\,,
\label{z1}
\end{equation}
is the familiar classical action. We implement fermion reparametrisation (\ref{rep}) by explicitly incorporating the axial phase field $\sigma$ into the generating functional:
\begin{equation}
Z_{\rm QED'}[V_{\mu}]=\int \mathcal{J}[\sigma, V_{\mu}]D\sigma D\bar{\chi}D\chi\, e^{iS_{\rm QED'}[\bar\chi,\chi, V_{\mu}, \sigma]}\,.
\label{z2}
\end{equation}
Here the ${\rm QED}'$ action reads:
\begin{equation}
S_{\rm QED'}[\bar\chi,\chi, V_{\mu}, \sigma]=S_{\rm QED}[\bar\chi,\chi, V_{\mu}]-\int d^4x\, \partial_{\mu}\sigma \bar\chi \gamma^{\mu}\gamma_5\chi\,.
\label{qed'}
\end{equation}
The path integral (\ref{z2}) contains the formal (renormalised) functional Jacobian factor $\mathcal{J}[\sigma, V_{\mu}]$ of transformations (\ref{rep}):
\begin{equation}
D\bar\psi D\psi=\mathcal{J}[\sigma, V_{\mu}]D\sigma D\bar\chi D\chi\,. 
    \label{jac}
\end{equation}
There is a well-known ambiguity in the definition of the renormalised Jacobian $\mathcal{J}[\sigma,V_\mu]$, that includes ambiguities in the choice of the regulator as well as counterterms. Consequently, the original theory is not \emph{a priori} guaranteed to be quantum mechanically equivalent to the theory formulated in terms of the reparametrised fields (in which case the formal Jacobian properly matches the change of variables after renormalisation). 
In particular, although the classical action $S_{\rm QED'}$ is manifestly invariant under the local axial transformation (\ref{gauge}), the renormalised Jacobian $\mathcal{J}[\sigma,V_\mu]$ need not preserve this symmetry for an arbitrary choice of its local counterterms. For example, even if one insists on preserving the vector gauge symmetry of ${\rm QED}$, as required for a consistent quantum theory, while allowing the axial reparametrisation symmetry to be anomalous, the resulting ${\rm QED}'$ theory need not be equivalent to ${\rm QED}$: the axial phase $\sigma$, which was a redundant variable at the classical level, may acquire physical dynamics and represent an additional degree of freedom in otherwise healthy effective field theory.

Conversely, if the renormalisation prescription is chosen so as to preserve both the vector gauge symmetry and the local axial reparametrisation symmetry, the $\sigma$-dependence becomes a pure gauge redundancy. That is, the $\sigma$ field can be removed by the appropriate gauge choice and the reparametrised theory is physically equivalent to the original ${\rm QED}$. 

To see this explicitly, we first note that invariance of the generating functional (\ref{z2}) under the local gauge transformations,
\begin{equation}
    \chi'=  e^{i\beta}\chi\,,\quad \bar\chi'=\bar\chi e^{-i\beta}\,,\quad V_\mu'=V_\mu+\partial_\mu \beta\,, \label{vector_gauge_transformation}
\end{equation}
implies $\partial_{\mu}\left (\frac{\delta Z_{\rm QED'}}{\delta V_{\mu}}\right) = 0$. This translates to the anomaly equation for the divergence of expectation value of the fermionic vector current, $j_V^{\mu}\equiv \left\langle\frac{\delta S_{\rm QED'}}{\delta V_{\mu}} \right\rangle=\left\langle \bar\chi\gamma^{\mu}\chi\right\rangle$ :
\begin{equation}   \partial_{\mu}j^{\mu}_V=i\partial_{\mu}\left\langle\frac{\delta \ln\mathcal{J}}{\delta V_{\mu}} \right\rangle\,.
\label{vanom}
\end{equation}

Next, we take the functional derivative of Eq.~(\ref{z2}) with respect to $\sigma$, to obtain the anomaly equation for the divergence of the fermion axial current, $j^{\mu}_A\equiv \left\langle\bar\chi \gamma^{\mu}\gamma^5\chi\right\rangle$: 
\begin{equation}
\partial_{\mu}j_A^{\mu}\equiv \left\langle \frac{\delta S_{\rm QED'}}{\delta \sigma}\right\rangle = -i\left\langle \frac{\delta\ln\mathcal{J}}{\delta \sigma}\right\rangle\,.
\label{canom}
\end{equation}

If the original (\ref{QED_Z}) and reparametrised (\ref{z2}) generating functionals are required to describe the same physical theory, the following constraint must be satisfied:
\begin{equation}
\int D\sigma \mathcal{J}[\sigma,V_\mu]
{\rm e}^{i\big(S_{QED'}[\bar\chi,\chi,V_{\mu}]-S_{QED}[\bar\chi,\chi,V_{\mu}]\big)}=1\,.
\label{cond}
\end{equation}
This constraint must hold in any gauge and hence remain invariant under local gauge transformations (\ref{vector_gauge_transformation}). This requirement imposes the following condition on the anomalous Jacobian:
\begin{equation}
\partial_{\mu}\left(\frac{\delta\ln\mathcal{J}}{\delta V_{\mu}}\right)=0\,.
\end{equation}
This condition is simply the gauge invariance of the Jacobian; compare this to the situation in Eq.~(\ref{canom}), where the anomalous variation of the Jacobian must cancel with the variation of the action to ensure overall reparametrisation invariance.
Consequently, imposing both local vector gauge invariance and axial reparametrisation invariance leads to the familiar Adler–Bell–Jackiw (ABJ) anomaly:
\begin{equation}  \partial_{\mu}j^{\mu}_V=0\,,\quad\partial_{\mu}j^{\mu}_A=-i\left\langle \frac{\delta\ln\mathcal{J}}{\delta \sigma}\right\rangle\,.
\label{abj}
\end{equation}

We emphasise that, from the perspective of the effective theory, vector gauge invariance is a necessary consistency requirement, whereas reparametrisation invariance is not directly tied to the consistency of the theory. In fact, violation of the condition (\ref{cond}) signals the presence of additional degrees of freedom, ultimately associated with the ultraviolet completion of the effective theory. In the remainder of this section, we identify perturbative anomalies as a source of violations of the reparametrisation-invariance condition (\ref{cond}). The purely non-perturbative mechanism for violating this condition, together with the emergence of the axial field $\sigma$ as a propagating degree of freedom, will be discussed in the following section.

Working within a diagrammatic formalism, there are four distinct triangle diagrams of a fermion loop that contain vector (V) and axial (A) vertices: VVV, AVV, VAA and AAA. In a vector-like theory with Dirac fermions, VVV diagrams sum to zero. The standard AVV diagrams with cutoff regularisation and generic momentum routing parameters\footnote{The momentum-routing ambiguity is a manifestation of a fundamental ambiguity inherent in regularisation schemes. For example, in dimensional regularisation, although amplitudes are independent of the choice of momentum routing, an analogous ambiguity reappears in the definition and treatment of chiral objects in $(4-\epsilon)$ dimensions.} $(c_1,c_2)$ (see Appendix \ref{app:momentum_routing} for our conventions) produce the following anomalous divergences in vector (\,$j^{\mu}_V=\langle \bar\psi\gamma^{\mu}\psi\rangle $) and axial (\,$j^{\mu}_A$) currents, respectively: 
\begin{align}
\partial_{\mu}j^{\mu}_V&\supset \frac{c_2-c_1-2}{8\pi^2}\epsilon^{\mu\nu\alpha\beta} \partial_\mu A_\nu \partial_\alpha V_\beta
\label{ABJ_vector_anomaly}\\
\partial_{\mu}j^{\mu}_A&\supset \frac{c_1-c_2}{8\pi^2}\epsilon^{\mu\nu\alpha\beta} \partial_\mu V_\nu \partial_\alpha V_\beta\,.
\label{ABJ_axial_anomaly}
\end{align}
One normally chooses the routing parameters\footnote{A specific choice of the momentum routing parameters corresponds to a specific choice of counterterms.} such that the vector current is divergence-free: $c_2-c_1=2$, and hence one is left with an anomalous axial current, as in Eq.~(\ref{ABJ_axial_anomaly}). Note that despite the anomalous divergence of the current, the axial charge, $Q_A=\int d^3x\langle\hspace{1pt} j^0_A\rangle $, remains conserved for any physical in-vacuum process.\footnote{This picture is known to be dramatically altered in non-Abelian gauge theories, but also in Abelian theories with non-trivial topology; see Refs.~\cite{Deser:1980kc, Isham:1983qy,  Kobakhidze:2026pwp, Arunasalam:2018eaz} and the discussion below.}
  
Note, however, that if one restricts to only longitudinal external axial vector fields, i.e., $A_{\mu}=\partial_{\mu}\sigma$, as this is the case in our reparametrised theory (\ref{z2}), the divergence of the vector current (\ref{ABJ_vector_anomaly}) vanishes automatically. Therefore, it seems there is an arbitrariness in selecting the routing parameters, and hence the counterterm in the effective action:
\begin{equation}
    \mathcal{L}_{\rm eff} \supset \frac{(2-\kappa)\sigma}{16\pi^2}F_{\mu\nu}\tilde{ F}^{\mu\nu}\,,
\end{equation}

If we do not require the physical currents to be reparametrisation-invariant, the additional contribution to the coefficient of the axial anomaly (\ref{ABJ_axial_anomaly}), $\kappa = 2-c_1+c_2$, can be interpreted as arising from additional heavy fermions. This contribution is uniquely determined once the full fermion spectrum of the ultraviolet completion is specified, fixing the axion-like coupling between $\sigma$ and the vector gauge field. 

At the same time, the vector current inferred from Eq.~(\ref{ABJ_vector_anomaly}) is: $j^{\mu}_V=\kappa\partial_{\nu}\sigma\tilde F^{\mu\nu}$. Although this current is manifestly gauge-invariant, it is not invariant under the local shift of $\sigma$ in Eq.~(\ref{gauge}) unless $\kappa=0$, which corresponds to the standard choice of routing parameters, $c_2-c_1=2$. Thus, requiring the physical observables to be reparametrisation-invariant fixes the anomaly coefficient to its standard ABJ value, arising entirely from the AVV triangle diagram involving the low-energy Dirac fermion only.      

Let us consider now the AAA diagrams. They produce the following contribution to the axial anomaly:
\begin{equation}
\partial_{\mu}j^{\mu}_A\supset\frac{c_3-c_4}{8\pi^2}\epsilon^{\mu\nu\alpha\beta} \partial_\mu A_\nu \partial_\alpha A_\beta\,,   
\end{equation}
where again $(c_3,c_4)$ are the routing parameters. However, this time the Bose symmetry of AAA diagrams fixes them to be $c_3=-c_4=-1/3$. In ${\rm QED}'$, $A_{\mu}=\partial_{\mu}\sigma$, and hence this anomaly contribution vanishes.  

Finally, consider VAA diagrams. They produce the following contributions to the vector and axial anomalies: 
\begin{align}
\partial_{\mu}j^{\mu}_V&\supset -\frac{c_5+c_6}{24\pi^2}\eta^{\mu\nu\alpha\beta}\, \partial_\mu(A_\alpha \partial_\beta A_\nu) \\
\partial_{\mu}j^{\mu}_A&\supset \frac{c_5+c_6}{24\pi^2}\eta^{\mu\nu\alpha\beta}\, \partial_\mu(A_\alpha \partial_\beta V_\nu)\,,    
\end{align}
where $\eta^{\alpha\beta\gamma\delta}=\eta^{\alpha\beta}\eta^{\gamma\rho}+\eta^{\alpha\gamma}\eta^{\beta\rho}+\eta^{\alpha\rho}\eta^{\beta\gamma}$ is a totally symmetric Lorentz-invariant tensor.
Both of these anomalies can be removed by choosing routing parameters as, $c_5=-c_6$. These are known as irrelevant anomalies. However, allowing $c_5\neq -c_6$, we observe that the anomalies persist in ${\rm QED}$ even for vacuum configurations, i.e., $A_{\mu}=\partial_{\mu}\sigma,~V_{\mu}=\partial_{\mu}\rho$. To reproduce these anomalies we may equivalently add the counterterm in the effective Lagrangian\footnote{In the BRST language, the fact that irrelevant anomalies may be completely reproduced with a counterterm means that the anomaly is \textit{BRST exact} so falls in the the trivial cohomology class \cite{Barnich:2000}.}:
\begin{equation}
    {\mathcal L}_{\rm eff} \supset \frac{C}{48\pi^2} \eta^{\mu\nu\alpha\beta}\partial_\mu (\partial_\nu\sigma \partial_\alpha\sigma)V_{\beta}\,,
\label{ct}
\end{equation}
where $C=c_5+c_6$ and we have set $A_{\mu}=\partial_{\mu}\sigma$.  
Remarkably, the theory remains gauge invariant under the ${\rm QED}$ and axial reparametrisation transformations provided the vector and axial current anomalies hold:
\begin{align}
\label{irrelevantanomalyequations}
\partial_{\mu}j^{\mu}_V&=-\frac{C}{24\pi^2}\eta^{\mu\nu\alpha\beta} \partial_\mu(\partial_\alpha\sigma \partial_\beta \partial_\nu\sigma ) \\
\partial_{\mu}j^{\mu}_A&=\frac{C}{24\pi^2}\eta^{\mu\nu\alpha\beta} \partial_\mu(\partial_\alpha\sigma \partial_\beta V_\nu )+\frac{1}{8\pi^2}F_{\mu\nu}\tilde F^{\mu\nu}\,.
    \label{va}
\end{align}
We now observe that imposing $U(1)$ gauge invariance together with local axial reparametrisation invariance, which enforces an exact matching of the numbers of degrees of freedom between the classical and quantum theories, renders the parameter $C$ redundant, since neither the gauge-invariant nor the axial-reparametrisation observables depend on it. In other words, the anomaly can be "absorbed" in the definition of physical currents: $j^{\mu}_V +\frac{C}{24\pi^2}\eta^{\mu\nu\alpha\beta} \left(\partial_\alpha\sigma \partial_\beta \partial_\nu\sigma \right)$ and $j^{\mu}_A-\frac{C}{24\pi^2}\eta^{\mu\nu\alpha\beta} \left(\partial_\alpha\sigma \partial_\beta V_\nu \right)$, which are gauge and reparametrisation invariant objects. This completely fixes the regularisation ambiguity, leaving only the standard ABJ anomaly.

On the other hand, the vector anomaly equation (\ref{irrelevantanomalyequations}) is gauge invariant, and hence $j^\mu_V$ itself is a gauge-invariant observable. Therefore, if we relax axial reparametrisation invariance, the counterterm (\ref{ct}) can contribute to physical processes, resulting in anomalous non-conservation of the corresponding fermionic current. The precise form and physical interpretation of this non-conservation depend on the ultraviolet completion of the theory, which must contain additional degrees of freedom. In this case, the original and reparametrised theories are not equivalent.

Concluding this section, we would like to emphasise an important distinction between the coefficient $C$ of the irrelevant anomaly and the anomaly coefficient of the ABJ anomaly. The latter is famously not renormalised \cite{Adler:1969er}, owing to the topological nature of the corresponding operator. In contrast, the operator in Eq.~(\ref{ct}) is not topological and therefore does not enjoy the same cohomological protection. Consequently, its coefficient receives higher-loop corrections, making $C$ a renormalisation-scale-dependent running parameter\footnote{This running does not cause issues with satisfying anomaly equations, since any matrix element of physical observable operators are renormalisation scale-independent. To maintain gauge invariance the cancellation between the counterterm (\ref{ct}) and the anomalous divergence (\ref{va}) should match scale by scale.}.  From this perspective, different values of $C$ associated with the fact that the corresponding routing parameters are not fixed, represent $C$ at different renormalisation scales. 

Furthermore, the operator in Eq.~(\ref{ct}) is irrelevant in the infrared, implying that $C\to 0$ in the infrared regime of the theory. Thus, theories with a physical irrelevant anomaly and theories in which this anomaly is absent both approach classical electromagnetism in the long-distance limit. 

\section{QED with non-trivial topology and an emergent axion}
\label{sec3}

In the previous section, we integrated in the generating functional the standard perturbative axial anomaly by introducing an auxiliary pseudoscalar field $\sigma$ and argued that it remains non-dynamical as long as the theory is invariant under local axial reparametrisations of the fermion fields. In this section, we explore a possible non-perturbative mechanism through which the $\sigma$ field can become a propagating degree of freedom.

The non-perturbative mechanism we have in mind arises in the presence of topologically non-trivial gauge fields. Since, in ${\rm QED}$ defined on a topologically trivial flat spacetime, the gauge fields themselves do not support non-trivial topological sectors, there are two possible ways to proceed. The first, and more conventional, possibility is to assume that the Abelian gauge group of ${\rm QED}$ is embedded into a compact gauge group above the cutoff scale $\Lambda$. Here, however, we focus on a second, less orthodox possibility, in which non-trivial topological features of the electromagnetic field are induced by the topology of the underlying spacetime manifold \cite{Deser:1980kc, Isham:1983qy,  Kobakhidze:2026pwp, Arunasalam:2018eaz}.\footnote{A non-Abelian embedding may be understood as a change in the effective spacetime in the infrared. For example, in the theory where an $SU(2)$ gauge theory with adjoint Higgs is spontaneously broken to $U(1)$, the existence of an 't Hooft--Polyakov monopole may be described in the infrared by cutting out a 't Hooft line from spacetime (with appropriate boundary conditions). The resulting effective spacetime $\mathbb{R}^4\backslash \mathbb{R}\cong S^2\times \mathbb{R}^2$ has the appropriate topology, $H^2(S^2\times \mathbb{R}^2)\cong \mathbb{Z}$ to realise the effects of the monopole in the UV theory.}

In particular, on compact manifolds ($M$)\footnote{These include manifolds which may be treated as compact due to an appropriate choice of boundary conditions, as in the usual treatment of instantons in Yang--Mills theory. See, for example, Ref.~\cite{Kobakhidze:2026pwp} for an example in Abelian gauge theory on a curved background.} with non-trivial second cohomology, $H^2(M)\cong H_2(M)\neq 0$, there exist non-contractible 2-cycles on which Abelian gauge fields can support topologically non-trivial configurations. These configurations are characterised by quantised fluxes of the Abelian field strength through the corresponding closed Gaussian surfaces $\Sigma$:
\begin{equation}
\frac{1}{2\pi}\int_{\Sigma} F
\in \mathbb{Z}\,.
\end{equation}
Furthermore, the existence of such configurations is equivalent to the existence of field strengths with non-zero integer Pontryagin number
\begin{equation}
    \frac{1}{8\pi^2}\int_M F\wedge F\in \mathbb{Z}\,.
\end{equation}

This additional topological structure has significant implications for the vacuum\footnote{The “vacuum” is understood here as being defined with respect to a fixed spacetime manifold. For example, in the case of a Euclidean black-hole background, it would correspond to the equilibrium thermal state defined on that fixed geometry which has smallest free energy \cite{Hartle:1976tp}.} structure of the theory. Most notably, the integrated anomaly equation implies the existence of normalisable zero modes of the Dirac operator. This is a direct manifestation of the Atiyah--Singer index theorem \cite{Atiyah:1963zz}.

Consequently, non-perturbative processes that violate axial charge, $
\frac{dQ_A}{dt}\neq 0,
$ become possible through the overlap of physical fermionic states with the vacuum zero modes, i.e. through the phenomenon of level crossing. In the simplest theory with a single massless Dirac fermion, the associated fermion zero modes generate a fermion mass term represented by the non-perturbative 't Hooft vertex. Furthermore, the zero modes give rise to a non-vanishing fermion condensate,\footnote{This can be seen as the local limit of the 't Hooft fermion correlator \cite{t_Hooft:1976}.}
\begin{equation}
\left\langle \bar\psi\psi \right\rangle\neq0\,,
\end{equation}
which breaks the axial symmetry spontaneously, in addition to its explicit breaking by a fermion mass term. The collective axial phase of this condensate, previously represented by the non-dynamical pseudoscalar field $\sigma$, can then acquire a kinetic term and emerge as a propagating degree of freedom. In this sense, the $\sigma$ field is promoted from an auxiliary field to a genuine infrared degree of freedom by the non-perturbative dynamics associated with the topologically non-trivial sector.

To see this explicitly, we inspect the anomalous Ward identity:
\begin{equation}
    \bigg\langle \bigg(\int F\wedge F\bigg)\, \bar\psi\gamma_5\psi \bigg\rangle \propto\left\langle \bar \psi\psi\right\rangle\,, 
\end{equation}
found by varying the $\langle\bar\psi\gamma_5\psi\rangle$ with respect to a local axial transformation.
By inserting a complete set of states, the non-vanishing condensate on the RHS of the above equation implies the existence of a particle state $\vert \sigma \rangle$ that is annihilated/created into/from the vacuum state by local composite operators $F_{\mu\nu}\tilde F^{\mu\nu}$ and $\bar\psi\gamma_5\psi$, that is:
\begin{equation}
\left\langle {\rm vacuum}\right \vert F_{\mu\nu}\tilde F^{\mu\nu}\left\vert \sigma\right\rangle \neq 0\,,\quad\left\langle \sigma \right\vert \bar\psi\gamma_5\psi\left\vert  {\rm vacuum}\right\rangle \neq 0\,.
\end{equation}

This is a version of the Goldstone theorem for anomalous symmetry. As well-established by now \cite{Aurilia:1980jz, Dvali:2005an, Dvali:2005ws}, spontaneous breaking of anomalous symmetry is realised in the Higgs regime for the composite 3-form gauge field $C_3$, where $dC_3=F\wedge F$. Modulo topological subtleties described in Ref.~\cite{Canete:2026vxk}, the massive 3-form field is dual to a massive propagating pseudoscalar field $\sigma$. See also Refs.~\cite{Sakhelashvili:2021eid, Dvali:2024dlb, Dvali:2024zpc, Dvali:2025pcx} describing the same universal phenomenon in different contexts.

Hence, non-trivial topology inherited in the Abelian gauge field either through its embedding in non-Abelian gauge theory in the ultraviolet or through topologically non-trivial manifold leads to the prediction of additional pseudoscalar particle state in the theory.

\section{Conclusion and discussion}
In this paper, we have discussed various aspects of massless electrodynamics associated with the infrared behaviour of the theory. It is well known that, when treated as an effective theory, massless asymptotic fermion states lack a conventional particle interpretation, while the theory exhibits a quantisation ambiguity manifested through anomalies. We argue that this ambiguity can be formally fixed by requiring invariance under axial reparametrisations of the fermion fields, whereby global axial transformations are promoted to a gauge redundancy carried by the non-dynamical pseudoscalar field $\sigma$. In particular, reparametrisation invariance removes the so-called irrelevant anomalies, leaving only the standard ABJ axial anomaly, which is an intrinsically infrared phenomenon and is independent of the ultraviolet completion of the theory.

Alternatively, if reparametrisation invariance is not imposed, the irrelevant anomalies enter the effective theory through additional counterterms. Although the effective theory flows to standard electrodynamics in the infrared, the quantisation ambiguity associated with the regularisation and renormalisation procedure indicates that the theory retains sensitivity to ultraviolet physics and, in particular, may require additional degrees of freedom in its ultraviolet completion.

We have also discussed the fate of axial reparametrisation invariance in the presence of non-perturbative effects associated with non-trivial topology of the Abelian gauge field. Such configurations may arise when the theory is embedded into a non-Abelian gauge theory in the ultraviolet or when it is defined on a manifold with non-trivial topology. In this case, both the vacuum structure of the theory and its infrared particle spectrum can be dramatically modified.

We have argued that the axial anomaly in the presence of a non-zero topological Pontryagin charge supports in-vacuum fermion zero modes. These zero modes induce non-perturbative processes that violate axial charge conservation and are described by the corresponding ’t Hooft vertices. In the case of a single Dirac fermion, this generates a non-perturbative fermion mass.

Moreover, the zero modes are manifested through vacuum condensates of composite fermion operators, leading to spontaneous axial symmetry breaking. This breaking is realised in the Higgs regime and manifests itself through the emergence of a massive axion field. The latter can be viewed as the dynamical manifestation of the pseudoscalar $\sigma$, which was originally a non-dynamical redundant degree of freedom associated with axial reparametrisation invariance.

It would be interesting to extend the above analysis to another anomalous symmetry manifest in massless electrodynamics, namely scale invariance. In fact, the non-zero fermion condensate spontaneously breaks scale invariance as well. We therefore anticipate that, on manifolds with non-trivial Euler characteristic, the Higgsing of the gravitational composite 3-form may occur, giving rise to an emergent dilaton state. A related phenomenon in the context of supergravity was discussed in Ref.~\cite{Dvali:2024dlb}, where axial $R$-symmetry, scale invariance, and global supersymmetry are spontaneously broken by a gravitino condensate induced by gravitational instantons.

Finally, while we have discussed a prototype model of massless electrodynamics, the phenomena described here may naturally be  incorporated in realistic theories of particle physics and gravity. The infrared emergent states reflecting the topological features of such theories may have wide reaching phenomenological implications, see e.g., recent discussion in Ref.~\cite{Cesca:2026jua}.   

\section*{Acknowledgments}
We are grateful to Gia Dvali and Oto Sakhelashvili for valuable discussions on various topics related to this paper. This work was supported by the Australian Research Council through Discovery Project Grant DP220101721. JC and EL also acknowledge support from the University of Sydney through the Grants-in-Aid and PRSS travel grants.

\appendix

\section{Momentum routing in triangle diagrams}\label{app:momentum_routing}
\begin{figure}
    \centering
    \begin{tikzpicture}
        \begin{feynman}[small]
        \vertex (x);
        \vertex[right=1.5cm of x] (d1);
        \vertex[above=0.8cm of d1] (y1) ;
        \vertex[below=0.8cm of d1] (y2) ;
        \vertex[left=1cm of x] (ext0) {$V$};
        \vertex[right=0.75cm of y1] (d2);
        \vertex[above=0.9cm of d2] (ext1) {$A$};
        \vertex[right=0.75cm of y2] (d3);
        \vertex[below=0.9cm of d3] (ext2) {$A$};
        
            \diagram* {
                (x) -- [anti fermion] (y1);
                (ext1) -- [photon, momentum=$p$] (y1),
                (ext2) -- [photon, momentum=$q$] (y2),
                (y1) -- [anti fermion, edge label=$k+c_5p+c_6q$] (y2),
                (y2) -- [anti fermion] (x);
                (ext0) -- [photon, momentum=$-p-q$] (x);
            };
        \end{feynman}
    \end{tikzpicture}\qquad\begin{tikzpicture}
        \begin{feynman}[small]
        \vertex (x);
        \vertex[right=1.5cm of x] (d1);
        \vertex[above=0.8cm of d1] (y1) ;
        \vertex[below=0.8cm of d1] (y2) ;
        \vertex[left=1cm of x] (ext0) {$V$};
        \vertex[right=0.75cm of y1] (d2);
        \vertex[above=0.9cm of d2] (ext1) {$A$};
        \vertex[right=0.75cm of y2] (d3);
        \vertex[below=0.9cm of d3] (ext2) {$A$};
        
            \diagram* {
                (x) -- [fermion] (y1);
                (ext1) -- [photon, momentum=$p$] (y1),
                (ext2) -- [photon, momentum=$q$] (y2),
                (y1) -- [fermion, edge label=$k+c_6p+c_5q$] (y2),
                (y2) -- [fermion] (x);
                (ext0) -- [photon, momentum=$-p-q$] (x);
            };
        \end{feynman}
    \end{tikzpicture}
    \caption{Momentum routing in the VAA triangle diagrams. Bose symmetry of the vertex function is satisfied for any $(c_5, c_6)$. Note that the coefficients are swapped in the second diagram, where the fermion loop is reversed.}
    \label{fig:VAA}
\end{figure}
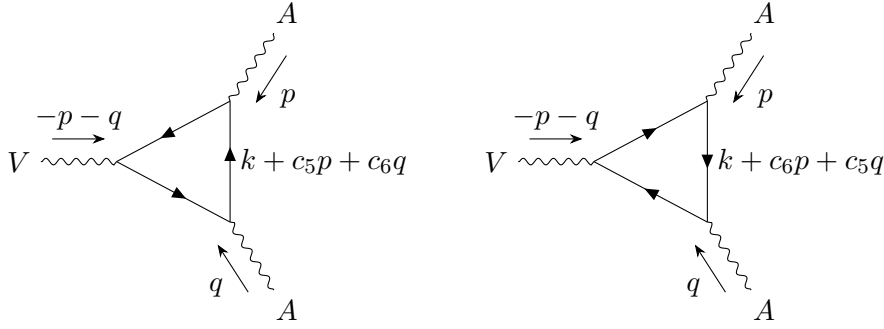

\begin{figure}[h]
    \centering
    \begin{tikzpicture}
        \begin{feynman}[small]
        \vertex (x);
        \vertex[right=1.5cm of x] (d1);
        \vertex[above=0.8cm of d1] (y1) ;
        \vertex[below=0.8cm of d1] (y2) ;
        \vertex[left=1cm of x] (ext0) {$A$};
        \vertex[right=0.75cm of y1] (d2);
        \vertex[above=0.9cm of d2] (ext1) {$V$};
        \vertex[right=0.75cm of y2] (d3);
        \vertex[below=0.9cm of d3] (ext2) {$V$};
        
            \diagram* {
                (x) -- [anti fermion] (y1);
                (ext1) -- [photon, momentum=$p$] (y1),
                (ext2) -- [photon, momentum=$q$] (y2),
                (y1) -- [anti fermion, edge label=$k+c_1p+c_2q$] (y2),
                (y2) -- [anti fermion] (x);
                (ext0) -- [photon, momentum=$-p-q$] (x);
            };
        \end{feynman}
    \end{tikzpicture}\qquad\begin{tikzpicture}
        \begin{feynman}[small]
        \vertex (x);
        \vertex[right=1.5cm of x] (d1);
        \vertex[above=0.8cm of d1] (y1) ;
        \vertex[below=0.8cm of d1] (y2) ;
        \vertex[left=1cm of x] (ext0) {$A$};
        \vertex[right=0.75cm of y1] (d2);
        \vertex[above=0.9cm of d2] (ext1) {$V$};
        \vertex[right=0.75cm of y2] (d3);
        \vertex[below=0.9cm of d3] (ext2) {$V$};
        
            \diagram* {
                (x) -- [fermion] (y1);
                (ext1) -- [photon, momentum=$p$] (y1),
                (ext2) -- [photon, momentum=$q$] (y2),
                (y1) -- [fermion, edge label=$k+c_2p+c_1q$] (y2),
                (y2) -- [fermion] (x);
                (ext0) -- [photon, momentum=$-p-q$] (x);
            };
        \end{feynman}
    \end{tikzpicture}
    \caption{Momentum routing in the AVV triangle diagrams. Bose symmetry of the vertex function is satisfied for any $(c_1, c_2)$.}
    \label{fig:AVV}
\end{figure}
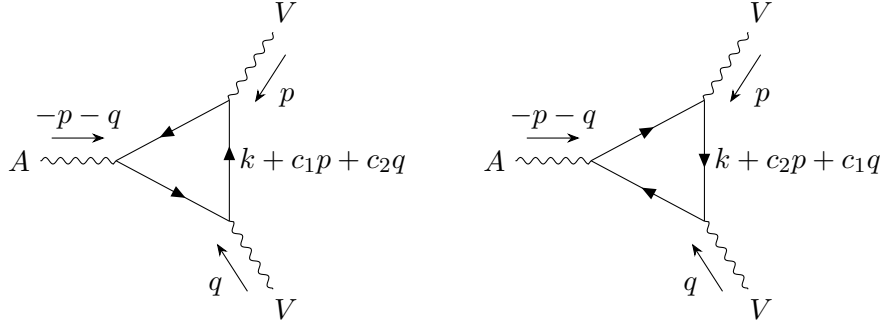

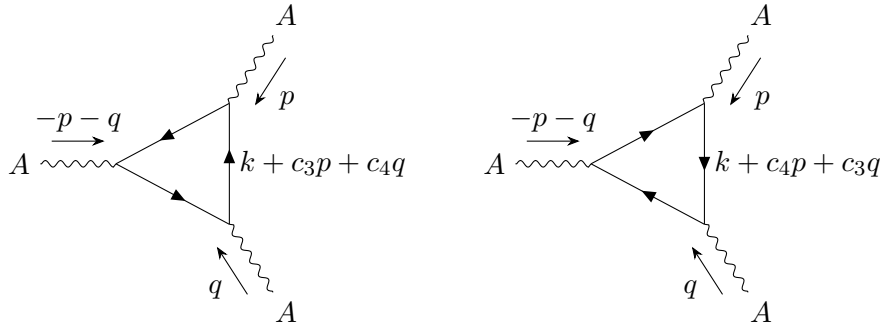
\begin{figure}[h]
    \centering
    \begin{tikzpicture}
        \begin{feynman}[small]
        \vertex (x);
        \vertex[right=1.5cm of x] (d1);
        \vertex[above=0.8cm of d1] (y1) ;
        \vertex[below=0.8cm of d1] (y2) ;
        \vertex[left=1cm of x] (ext0) {$A$};
        \vertex[right=0.75cm of y1] (d2);
        \vertex[above=0.9cm of d2] (ext1) {$A$};
        \vertex[right=0.75cm of y2] (d3);
        \vertex[below=0.9cm of d3] (ext2) {$A$};
        
            \diagram* {
                (x) -- [anti fermion] (y1);
                (ext1) -- [photon, momentum=$p$] (y1),
                (ext2) -- [photon, momentum=$q$] (y2),
                (y1) -- [anti fermion, edge label=$k+c_3p+c_4q$] (y2),
                (y2) -- [anti fermion] (x);
                (ext0) -- [photon, momentum=$-p-q$] (x);
            };
        \end{feynman}
    \end{tikzpicture}\qquad\begin{tikzpicture}
        \begin{feynman}[small]
        \vertex (x);
        \vertex[right=1.5cm of x] (d1);
        \vertex[above=0.8cm of d1] (y1) ;
        \vertex[below=0.8cm of d1] (y2) ;
        \vertex[left=1cm of x] (ext0) {$A$};
        \vertex[right=0.75cm of y1] (d2);
        \vertex[above=0.9cm of d2] (ext1) {$A$};
        \vertex[right=0.75cm of y2] (d3);
        \vertex[below=0.9cm of d3] (ext2) {$A$};
        
            \diagram* {
                (x) -- [fermion] (y1);
                (ext1) -- [photon, momentum=$p$] (y1),
                (ext2) -- [photon, momentum=$q$] (y2),
                (y1) -- [fermion, edge label=$k+c_4p+c_3q$] (y2),
                (y2) -- [fermion] (x);
                (ext0) -- [photon, momentum=$-p-q$] (x);
            };
        \end{feynman}
    \end{tikzpicture}
    \caption{Momentum routing in the AAA triangle diagrams. Bose symmetry of the vertex function is satisfied only for $(c_3, c_4) = (-\frac{1}{3},\frac{1}{3})$.}
    \label{fig:AAA}
\end{figure}

In this appendix we show the momentum routing and cutoff regularisation scheme producing the results quoted throughout the paper. As an example we will consider the VAA diagram. It's contribution to the effective action is given by 
\begin{equation}
    \Gamma_{VAA}^{(3)}[V,A] = \int\frac{d^4r\,d^4p\,d^4q}{(2\pi)^{12}} A_\nu(r)V_\alpha(p)V_\beta(q) \frac{1}{2!} \Gamma_{VAA}^{\nu\alpha\beta}(r,p,q)\,,
\end{equation}
in which the vertex function is formally
\begin{equation}
    \Gamma_{VAA}^{\nu\alpha\beta}(r,p,q) =(2\pi)^4\delta^{(4)}(r\!+\!p\!+\!q)\int \!\frac{d^4k}{(2\pi)^4}\!\left(\tr\left[\gamma^\nu \frac{i}{\slashed{k}\!+\!\slashed{p}} \gamma^5\gamma^\alpha \frac{i}{\slashed{k}} \gamma^5\gamma^\beta \frac{i}{\slashed{k}\!-\!\slashed{q}}\!+\!\gamma^\nu \frac{i}{\!-\!\slashed{k}\!+\!\slashed{q}} \gamma^5\gamma^\beta \frac{i}{\!-\!\slashed{k}} \gamma^5\gamma^\alpha \frac{i}{\!-\!\slashed{k}\!-\!\slashed{p}}\right]\right)\,,
\end{equation}
composed of fermions running clockwise and anticlockwise around the triangle loop. The loop momentum $k$ can be formally shifted by the incoming momentum as $k\rightarrow k+ c_5 p +c_6 q$ with $c_5$ and $c_6$ arbitrary momentum routing parameters as shown in Fig.~\ref{fig:VAA}. 
However, since the contribution has a superficial linear divergence, such shifts amount to a change in regularisation when we impose a fixed cutoff. The vector or axial-vector gauge variation of this contribution, which will contribute to the anomaly, depends on the contraction of the vertex function with the external momenta. For instance, the vector anomaly contribution is proportional to 
\begin{align}
    \frac{1}{2}ir_\mu \Gamma_{VAA}^{\nu\alpha\beta}(r,p,q) = \frac{1}{2} \delta(\cdots) \int \frac{d^4k}{(2\pi)^4} \bigg(&\tr\left[ \frac{1}{\slashed{k}+\slashed{s}-\slashed{q}}\slashed{r} \frac{1}{\slashed{k}+\slashed{s}+\slashed{p}}\gamma^\alpha \frac{1}{\slashed{k}+\slashed{s}}\gamma^\beta\right]\\
    &+\tr\left[ \frac{1}{\slashed{k}+\slashed{t}-\slashed{p}}\slashed{r} \frac{1}{\slashed{k}+\slashed{t}+\slashed{q}}\gamma^\beta \frac{1}{\slashed{k}+\slashed{t}}\gamma^\alpha\right]\bigg)\,,\nonumber
\end{align}
where $s= c_5p+c_6q$ and $t=c_6p+c_5q$, $(\gamma^5)^2=1$ has been used, and we have abbreviate the momentum conservation $\delta(\cdots)\equiv (2\pi)^4\delta^{(4)}(r+p+q)$ for brevity. The relative shifts of the two contributions $s$ and $t$ are chosen so as to preserve the Bose symmetry of the vertex function. Inserting $r = (k+s-q)-(k+s+p) = (k+t-p)-(k+t+q)$,
\begin{align}
    \frac{1}{2}ir_\mu \Gamma_{VAA}^{\nu\alpha\beta}(r,p,q) &= \frac{1}{2}\delta(\cdots) \int \frac{d^4k}{(2\pi)^4} \bigg(\tr\left[\frac{1}{\slashed{k}+\slashed{s}+\slashed{p}}\gamma^\alpha \frac{1}{\slashed{k}+\slashed{s}}\gamma^\beta\right]-\tr\left[ \frac{1}{\slashed{k}+\slashed{s}-\slashed{q}}\gamma^\alpha \frac{1}{\slashed{k}+\slashed{s}}\gamma^\beta\right]\\
    &\hspace{2.7cm}\qquad+\tr\left[\frac{1}{\slashed{k}+\slashed{t}+\slashed{q}}\gamma^\beta \frac{1}{\slashed{k}+\slashed{t}}\gamma^\alpha\right]-\tr\left[ \frac{1}{\slashed{k}+\slashed{t}-\slashed{p}}\gamma^\beta \frac{1}{\slashed{k}+\slashed{t}}\gamma^\alpha\right]\bigg)\nonumber\\
    &= \frac{1}{2}\delta(\cdots) \tr[\gamma^\mu\gamma^\alpha\gamma^\nu\gamma^\beta]\int \frac{d^4k}{(2\pi)^4} \bigg( \frac{(k+s+p)_\mu(k+s)_\nu}{(k+s+p)^2(k+s)^2}-\frac{(k+t)_\mu(k+t-p)_\nu}{(k+t)^2(k+t-p)^2}\\
    &\hspace{5cm}\qquad+ \frac{(k+t)_\mu(k+t+q)_\nu}{(k+t)^2(k+t+q)^2}-\frac{(k+s-q)_\mu(k+s)_\nu}{(k+s-q)^2(k+s)^2}\bigg)\,,\nonumber
    \intertext{which can be written in the form}
    &=2\delta(\cdots)(\eta^{\mu\alpha}\eta^{\nu\beta}-\eta^{\mu\nu}\eta^{\alpha\beta}+\eta^{\mu\beta}\eta^{\nu\alpha})\int \frac{d^4k}{(2\pi)^4} \Big(f_{\mu\nu}^{(1)}(k+l_1)-f_{\mu\nu}^{(1)}(k)\\
    &\hspace{8.5cm}+f_{\mu\nu}^{(2)}(k+l_2)-f_{\mu\nu}^{(2)}(k)\Big)\,,\nonumber
\end{align}
in which the two functions $f_{\mu\nu}^{(1)}(k) \equiv \frac{(k+t)_\mu(k+t-p)_\nu}{(k+t)^2(k+t-p)^2}$ and $f_{\mu\nu}^{(2)}(k) \equiv \frac{(k+s-q)_\mu(k+s)_\nu}{(k+s-q)^2(k+s)^2}$ and the two shifts in momentum are $l_1\equiv (c_5-c_6+1)p + (c_6-c_5)q$ and $l_2\equiv (c_6-c_5)p + (c_5-c_6+1)q$.\footnote{The calculation for the AVV and AAA anomalies look much the same but with a $\gamma^5$ appearing in the trace which produces the antisymmetric tensor $\epsilon^{\mu\nu\alpha\beta}$ instead of the metric structure.} Now, the integral over each function appears to be quadratically divergent; with appropriate regularisation, we can extract a finite surface term from their difference.  Taylor expanding the shifted functions, the quadratically divergent parts cancel,
\begin{align}
    \int \frac{d^4 k}{(2\pi)^4} [f_{\mu\nu}(k+l)-f_{\mu\nu}(k)] &= \int \frac{d^4 k}{(2\pi)^4} \left(l^\sigma \frac{d f_{\mu\nu}}{d k^\sigma} +\frac{1}{2}l^\sigma l^\lambda \frac{d^2 f_{\mu\nu}}{d k^\sigma dk^\lambda} +\cdots\right)\,,
    \intertext{where the higher order terms will go to zero at large $k$. We can write the integrand as a total derivative and so as a momentum-space boundary term,}
    \int \frac{d^4 k}{(2\pi)^4} [f_{\mu\nu}(k+l)-f_{\mu\nu}(k)]&= l^\sigma\int \frac{d^4 k}{(2\pi)^4}  \frac{d }{d k^\sigma}\left(f_{\mu\nu} +\frac{1}{2} l^\lambda \frac{d f_{\mu\nu}}{dk^\lambda} \right)\\
    &= l^\sigma \lim_{|k|\rightarrow \infty}\int \frac{d \Omega^{(3)}}{(2\pi)^4}  \hat{k}_\sigma |k|^3\left(f_{\mu\nu} +\frac{1}{2} l^\lambda \frac{d f_{\mu\nu}}{dk^\lambda} \right)\\
    &= l^\sigma \int \frac{d \Omega^{(3)}}{(2\pi)^4}   \lim_{|k|\rightarrow \infty}\hat{k}_\sigma|k|^3\left(f_{\mu\nu} +\frac{1}{2} l^\lambda \frac{d f_{\mu\nu}}{dk^\lambda} \right)\,,
\end{align}
where $\hat{k}_\sigma$ is a unit vector pointing outwards from the boundary. This is the cutoff regularisation; we temporarily assume a finite $|k|$ and then take the limit after extracting the finite result. For a function of the form $f_{\mu\nu} = \frac{(k+u)_\mu(k+v)_\nu}{(k+u)^2(k+v)^2}$, we can expand it and its derivative for large $|k|$,
\begin{align}
    \hat{k}_\sigma|k|^3 f_{\mu\nu}(k) &= |k|\hat{k}_\sigma\hat{k}_\mu\hat{k}_\nu + \hat{k}_\sigma\hat{k}_\nu u_\mu + \hat{k}_\sigma\hat{k}_\mu v_\nu - 2\hat{k}_\sigma\hat{k}_\mu\hat{k}_\nu\hat{k}_\lambda (u+v)^\lambda+\mathcal{O}(|k|^{-1})\\
    \hat{k}_\sigma|k|^3\frac{d f_{\mu\nu}}{dk^\lambda} &= \eta_{\mu\lambda}\hat{k}_\sigma\hat{k}_\nu+\eta_{\nu\lambda}\hat{k}_\sigma\hat{k}_\mu - 4\hat{k}_\sigma\hat{k}_\mu\hat{k}_\nu\hat{k}_\lambda+\mathcal{O}(|k|^{-1})\,.
\end{align}
Now we perform the angular integration. The linearly divergent pieces are odd in the unit vector $\hat{k}_\sigma$ and so vanish, the integral over a pair of unit vectors replaces them with $\hat{k}_\sigma\hat{k}_\mu \rightarrow \frac{\eta_{\sigma\mu}}{4}$ (the metric being the only symmetric rank 2 tensor available), and the integral over four unit vectors replaces them with $\hat{k}_\sigma\hat{k}_\mu\hat{k}_\nu\hat{k}_\lambda \rightarrow \frac{1}{24}(\eta_{\sigma\mu}\eta_{\nu\lambda}+\eta_{\sigma\lambda}\eta_{\nu\mu}+\eta_{\sigma\nu}\eta_{\mu\lambda})\equiv \frac{1}{24}\eta_{\sigma\mu\nu\lambda}$, ($\eta_{\sigma\mu\nu\lambda}$ being the only symmetric rank 4 tensor available). The surface integration also brings in an extra factor of the surface area of the unit 3-sphere $\int d\Omega^{(3)} = 2\pi^2$. Putting all of this together, 
\begin{align}
    \int \frac{d^4 k}{(2\pi)^4} [f_{\mu\nu}(k+l)-f_{\mu\nu}(k)]&= \frac{1}{8\pi^2} \bigg[l^\sigma\left(\frac{1}{4}\eta_{\sigma\nu}u_\mu+\frac{1}{4}\eta_{\sigma\mu}v_\nu-\frac{1}{12}\eta_{\sigma\mu\nu\lambda}(u+v)^\lambda\right)\\
    &\qquad\qquad\, + \frac{1}{2}l^\sigma l^\lambda \left(\frac{1}{4}\eta_{\mu\lambda}\eta_{\sigma\nu}+\frac{1}{4}\eta_{\mu\sigma}\eta_{\lambda\nu}-\frac{1}{6}\eta_{\sigma\mu\nu\lambda}\right)\bigg]\nonumber\\
    &= \frac{1}{96\pi^2}\left(l_\mu l_\nu +l_\nu(2u-v)_\mu+l_\mu (2v-u)_\nu-\eta_{\mu\nu} l_\lambda (u+v+l)^\lambda\right)\,.
\end{align}
We then contract this with $2(\eta^{\mu\alpha}\eta^{\nu\beta}-\eta^{\mu\nu}\eta^{\alpha\beta}+\eta^{\mu\beta}\eta^{\nu\alpha}) = 2(\eta^{\mu\alpha\nu\beta}-2\eta^{\mu\nu}\eta^{\alpha\beta})$. The terms proportional to just $\eta^{\alpha\beta}$ cancel, leaving only
\begin{align}
    2(\eta^{\mu\alpha\nu\beta}-2\eta^{\mu\nu}\eta^{\alpha\beta})\int \frac{d^4 k}{(2\pi)^4} [f_{\mu\nu}(k+l)-f_{\mu\nu}(k)] = \frac{1}{48\pi^2}\eta^{\mu\alpha\nu\beta} l_\mu(l+u+v)_\nu\,.
\end{align}
We now apply this to both pieces $f_{\mu\nu}^{(1)}(k)$, in which $l_1= (c_5-c_6+1)p + (c_6-c_5)q$,~ $u_1=c_6p+c_5q$, and  $v_1=(c_6-1)p+c_5q$ and $f_{\mu\nu}^{(2)}(k)$ in which $l_2= (c_6-c_5)p + (c_5-c_6+1)q$, ~$u_2=c_5p +(c_6-1)q$, and  $v_2=c_5p+c_6q$. We are left with the remarkably simple result,
\begin{align}
    \frac{1}{2}ir_\mu \Gamma_{VAA}^{\nu\alpha\beta}(r,p,q)=(2\pi)^4\delta^{(4)}(r+p+q)\frac{c_5+c_6}{48\pi^2}\eta^{\mu\alpha\nu\beta} (p+q)_\mu (p+q)_\nu\,.
\end{align}
Taking the Fourier transform, we get the final result for the vector anomaly contribution,
\begin{equation}
    \partial_\mu j^\mu_V \supset -\frac{c_5+c_6}{24\pi^2}\eta^{\mu\alpha\nu\beta} \partial_\mu(A_\alpha \partial_\beta A_\nu)\,.
\end{equation}
The calculations for the VAA contribution to the axial anomaly and the contributions from the AVV and AAA diagrams follow in much the same way. Their momentum routing is shown in Figs.~\ref{fig:AVV} and \ref{fig:AAA}, respectively. 
\vspace{2cm}
\vspace{1.5cm}

\newpage



\end{document}